\documentclass[a4paper]{article}
\usepackage{amsmath,amssymb,amsfonts,bm,xcolor}

\usepackage{algorithm,algorithmic,url}

\usepackage{enumitem,hyperref,natbib,multirow,rotating,sectsty,booktabs}
\allsectionsfont{\sffamily}
\usepackage[margin=1in]{geometry}

\usepackage{setspace}

\title{Sparse Bayesian ARX models\\ with flexible noise distributions}

\author{Johan Dahlin, Adrian~Wills and Brett~Ninness%
\thanks{E-mail adresses to authors: \url{firstname.lastname@newcastle.edu.au}. JD and AW are with the School of Engineering, The University of Newcastle, Australia. BN is with the Faculty of Engineering and Built Environment, The University of Newcastle, Australia. This work was supported by the Australian Research Council Discovery Project DP140104350.}%
}

\begin{document}
\maketitle

\doublespacing

\begin{abstract}
	\noindent
	This paper considers the problem of estimating linear dynamic system models when the observations are corrupted by random disturbances with nonstandard distributions.
	The paper is particularly motivated by applications where sensor imperfections involve significant contribution of outliers or \emph{wrap-around} issues resulting in multi-modal distributions such as commonly encountered in robotics applications.
	As will be illustrated, these nonstandard measurement errors can dramatically compromise the effectiveness of standard estimation methods, while a computational Bayesian approach developed here is demonstrated to be equally effective as standard methods in standard measurement noise scenarios, but dramatically more effective in nonstandard measurement noise distribution scenarios. \\

	\noindent \textbf{Keywords}:
	Bayesian inference, Hamiltonian Monte Carlo, Gaussian mixture models.
\end{abstract}

\newpage
\maketitle

\section{Introduction}
\label{sec:intro}
The distribution of the noise is an important assumption in modelling dynamical systems.
This as the accuracy of the parameter estimates of the model obtained by system identification methods \citep{Ljung1999} can suffer if the noise model is wrong or does not capture the main characteristics of the noise.
This is possibly a common problem as outliers and other un-modelled mechanisms naturally occur in many real world situations.
In this paper, we consider and illustrate the potentially dramatic effect of noise distribution mis-modelling and develop a method which estimates the noise model internally without detriment to system parameter estimates.

In particular, we consider a Bayesian approach~\citep{Robert2007,GelmanCarlinSternDunsonVehtariRubin2013} to dynamic system estimation.
We illustrate this approach using the very commonly employed auto-regressive exogenous input (ARX) model
\begin{align}
	A(q) y_t
	=
	B(q)
	u_t
	+
	e_t,
	\label{eq:arx}
\end{align}
where $y_t \in \mathbb{R}$ and $u_t \in \mathbb{R}$ denotes the output
and input of the system at time $t$ respectively and the system polynomials are given by
\begin{align}
	A(q) = 1 + \sum_{k=1}^{n_a} a_k q^{-k}, \quad
	B(q) = \sum_{k=1}^{n_b} b_k q^{-k},
	\label{eq:arx:polynomials}
\end{align}
where $a_{1:n_a} \triangleq \{a_k\}_{k=1}^{n_a}$ and $b_{1:n_b}$
denote the unknown system polynomial coefficients.
The shift operator is denoted by $q$, i.e., $q^{-k} y_t = y_{t-k}$.
Here, it is assumed that the system orders $n_a$ and $n_b$ are unknown parameters to be estimated from data.

To be able to handle a wide range of noise distributions, we assume that the noise $e_t$ can be modelled using a Gaussian mixture model (GMM) parameterised by
\begin{align}
	p(e_t)
	=
	\sum_{k=1}^{n_e}
	w_k\,
	\mathcal{N}(e_t; \mu_k, \sigma^2_k),
	\quad
	\sum_{k=1}^{n_e}
	w_k
	=
	1,
	\label{eq:noisemodel}
\end{align}
where $w_k$, $\mu_k$ and $\sigma_k$ denote the weight, mean and standard deviation of the mixture component $k$, respectively.
The GMM can successfully capture the behaviour of many types of commonly encountered
noise distributions including multi-model, skewed and long/heavy-tailed cases.

The main contributions of this paper are the following:
\begin{itemize}
	\item[(i)] developing a novel Bayesian ARX (BARX) model.
	\item[(ii)] tailoring an efficient sampling scheme for inference.
\end{itemize}

The main features of BARX is that:
\begin{itemize}
	\item[(a)] it includes sparsity promoting priors to automatically determine the required model orders $n_a$ and $n_b$ from data and;
	\item[(b)] the inclusion of a GMM to model the noise in a semi-parametric manner, which adapts its flexibility automatically to promote a sparse solution.
\end{itemize}

The inference is carried out using Hamiltonian Monte Carlo (HMC; \citealp{Neal2010,Betancourt2017}), which is an efficient Markov chain Monte Carlo (MCMC) scheme for estimating high dimensional posterior distributions.

We offer three numerical illustrations to study the BARX model using synthetic and real-world data.
These establish that the approach can perform well in different scenarios and that its data-driven nature is able to correctly estimate ARX model parameters, model orders and noise distributions from data.
Moreover, we observe that BARX can provide superior one-step-ahead predictors in comparison with other Bayesian approaches.

There is much related previous work to the present paper.
These include work within signal processing and system identification to allow for handling outliers within ARX models, see e.g., \cite{ChristmasEverson2011}, \cite{DahlinLindstenSchonWills2012} and \cite{TroughtonGodsill1998}.
However, the BARX model also allows for skewed and multi-modal noise, which is more general than just handling outliers.

Finite and infinite mixtures of Gaussian have a long history in statistics and machine learning, see e.g.,\ \cite{FruhwirthSchnatter2006} and \cite{EscobarWest1995}.
For example, \cite{MalsinerFruhwirthGrun2016} consider similar problems but makes use of Gibbs sampling and do not consider time series models, only regression.
Moreover, \cite{BaldacchinoWordenRowson2017} considers a mixture of Student's $t$ distributions for a class of time series models (not ARX) and use variational Bayes to estimate the posterior.


\section{Sparse Bayesian modelling}
\label{sec:modelling}
This paper employs a Bayesian approach to the estimation of the model orders, system parameters and noise mixture model coefficients in the model structure (\ref{eq:arx})-(\ref{eq:noisemodel}).
Delivering this rests on two key aspects - the selection of priors on all components to be estimated and provision of a means to combine these with the data likelihood to provide posterior estimates.  In this section we will address prior distribution selection.

\subsection{System polynomial coefficients}
The estimation of the system polynomials is basically a linear regression problem.
In the Bayesian setting, this requires us to choose a prior distribution for the coefficients.
The typical choice is a uniform distribution or a Gaussian distribution, which leads to a closed-form expression for the posterior if the likelihood is Gaussian.
These priors are known as conjugate priors in the Bayesian literature, see e.g., \cite{Bishop2006} and \cite{Murphy2012}.
Furthermore, the use of a Gaussian prior is equivalent to $L_2$-regularised least squares.

In this work, we assume a different prior that induces sparsity - unnecessary coefficients in an overparametrized model are shrunk towards zero.
This allows us to select a maximum model order for the system polynomials and the sparsity promoting mechanism will decide the appropriate model order from the data.

The optimal sparsity promoting prior should have a large probability mass at zero and long tails to accommodate for possible large system polynomial coefficients.
A prior with these qualities is given by
\begin{align}
	[a_{1:n_a} b_{1:n_b}]
	 & \sim
	\mathcal{N}(0, \sigma_f^2),
	\quad
	\sigma_f
	\sim
	\mathcal{C}_+(0, 1),
	\label{eq:horseshoe:ab}
\end{align}
where $\mathcal{C}_+(\mu, \gamma)$ denotes the Cauchy distribution restricted to be positive with location $\mu \in \mathbb{R}$ and scale $\gamma >0$.

This choice corresponds to a so-called \textit{horseshoe prior}, which exhibits the above desired properties since the Cauchy distribution puts a large fraction of its probability mass around its mode while also possessing infinite variance.
However, it has tails that decay geometrically and therefore coefficients will be shrunk towards zero if there is no evidence in the data saying otherwise.

The horseshoe prior has been shown to perform better (in the MSE sense) in regression problems than e.g., a Laplacian or Gaussian priors, see e.g., \cite{PolsonScott2010} and \cite{ArmaganClydeDunson2011}.
These benchmarks and promising results in a pilot study (not presented here) are the main reasons for this choice of prior distribution.

\begin{figure}[p]
	\centering
	\includegraphics[width=0.75\textwidth]{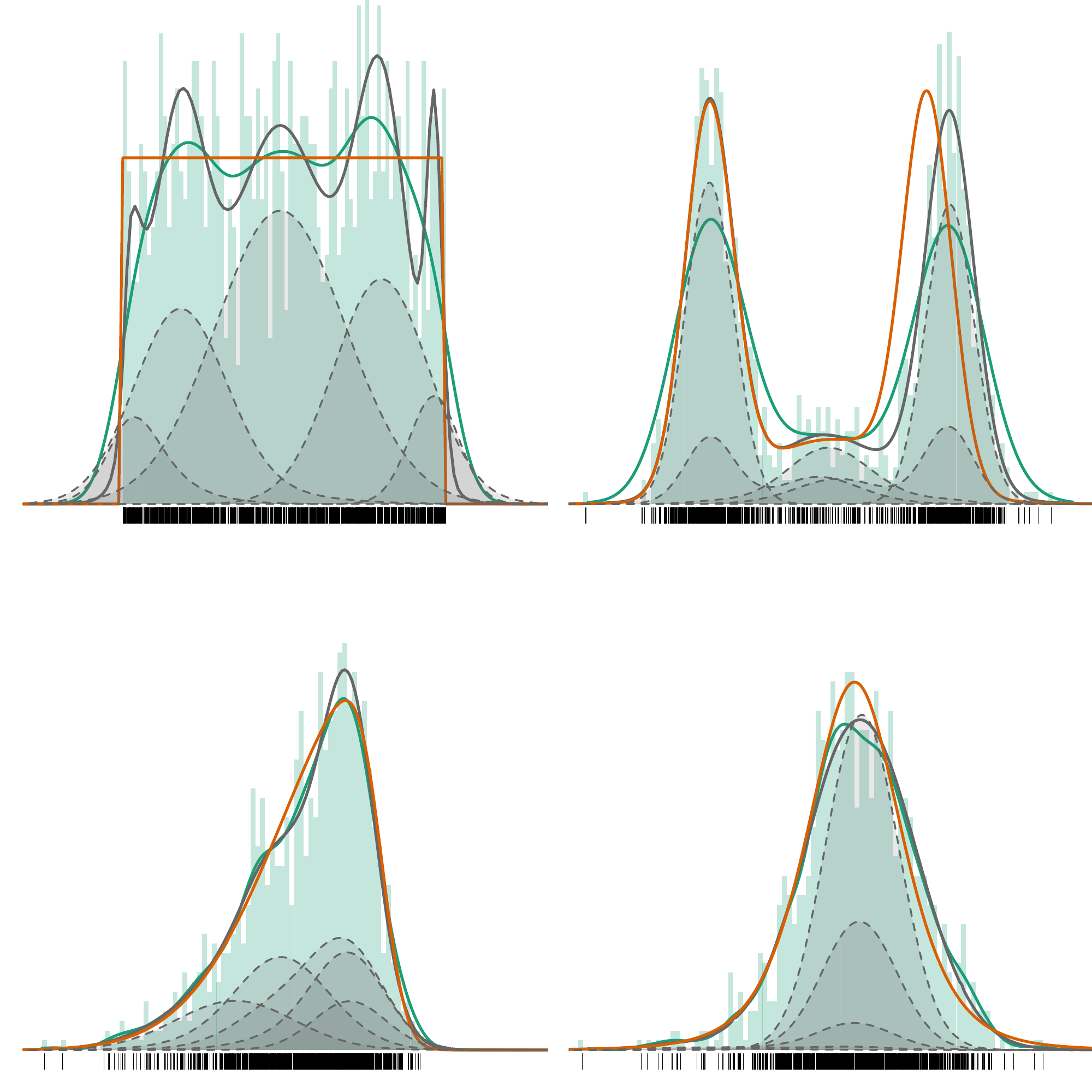}
	\caption{Four examples of GMMs (orange lines) for capturing the behaviour of the noise: uniform noise (top left), multi-modal noise (top right), skewed noise (bottom left) and heavy-tailed noise (bottom right).
		The histogram of the generated data is presented together with the kernel density estimate (green lines) and the GMM approximation (gray lines).
		The mixture components in the GMM are presented as gray lines and areas.}
	\label{fig:illustration-gmms}
\end{figure}

\subsection{Noise distribution}
The noise distribution is given by the GMM in \eqref{eq:noisemodel} with unknown order, weights, means and variances.
This is a powerful and versatile model class that covers many different interesting distributions as illustrated in Figure~\ref{fig:illustration-gmms}.

Here, the green histograms present the data simulated from four different distributions (orange lines): a uniform distribution, a GMM, a skewed Gaussian and a heavy-tailed Student's $t$.
The solid green line is the corresponding kernel density estimates formed from these histograms.
The GMM distribution components are shown in shaded grey with their summation as a solid grey line.

Note that in all these cases the GMM can successfully model all of these distributions using only five components.
As a consequence, the model structure (\ref{eq:arx})-(\ref{eq:noisemodel}) will be able to accommodate all these types of noise distributions without prejudicing the quality of the estimates of the system polynomials \eqref{eq:arx:polynomials}.

A popular approach for Bayesian inference in mixture models is to assume a non-parameteric prior known as the Dirichlet process (DP; \citealp{GelmanCarlinSternDunsonVehtariRubin2013,EscobarWest1995}) on the mixture weights.
The benefit of a DP is that it automatically determines the number of components required to describe the data and is flexible in the sense the this number increases with the number of observations.

However the drawback is that inference in such models can be challenging, especially using common MCMC methods as realisations from the associated Markov chain can exhibit poor mixing which results in a large computational burden.
Furthermore, the fact that the expected number of components grows with the amount of data is not desirable in our setting as the aim is to find the \textit{correct} number of components of the noise distribution.

Instead, we make use of recent progress in introducing sparsity in over-parameterised Bayesian finite mixture models.
This allows for more efficient inference and avoids the numerical challenges connected with infinite mixture models.
To achieve this, as in \eqref{eq:horseshoe:ab} we assume a sparsity inducing prior for the component means
\begin{align*}
	\mu_{1:n_e}
	 & \sim
	\mathcal{N}(0, \sigma_{\mu}^2),
	\quad
	\sigma_{\mu}
	\sim
	\mathcal{C}_+(0, 1),
\end{align*}
Moreover, we follow the recommendations by \cite{GelmanCarlinSternDunsonVehtariRubin2013} and also select a half-Cauchy distribution as the prior for $\sigma_{1:n_e}$,
\begin{align*}
	\sigma_{1:n_e}
	\sim
	\mathcal{C}_{+}(0, 5),
\end{align*}
which corresponds to a weakly informative prior on $\sigma_{1:n_e}$.

Finally, we assume a Dirichlet prior for the mixture weights
\begin{align*}
	w_{1:n_e}
	\sim
	\mathcal{D}(e_0, \ldots, e_0),
\end{align*}
where $n_e$ denotes the maximum number of components that could be present in the mixture.
The Dirichlet distribution is a standard choice in Bayesian mixture models as it is a distribution over the simplex, which means that the sum of $w_{1:n_e}$ is one at all times.
Note that, the Dirichlet distribution is the finite dimensional equivalent to the Dirichlet process.
Furthermore, it is also the appropriate conjugate prior for the the multinomial distribution, which allows for closed-form expressions for the posterior of the mixture weights.

The Dirichlet distribution can be used to introduce sparsity in the mixture by setting $e_0$ to a small number.
This results in only a few weights receiving the majority of the probability mass and therefore creating a mixture with a few active components if the data does not strongly support more active components.

A popular choice introduced by \cite{IshwaranZarepour2002} is to select $e_0 = 0.1 n_e^{-1}$ as the value of the hyperprior.
This empties superfluous components of the mixture and mimics the behaviour of the DP with concentration parameter $0.1$ asymptotically \citep{RousseauMengersen2011}.
However, empirical studies by e.g., \cite{MillerHarrison2013} have suggested that this approach typically over-estimate the number of components.
Another option is to use a hyper-prior for the concentration parameter as proposed by \cite{MalsinerFruhwirthGrun2016},
\begin{align}
 	e_0
 	\sim
 	\mathcal{G}
 	(\alpha_w, n_e \alpha_w),
 	\label{eq:sparse:weights:prior}
\end{align}
which is a Gamma distribution with mean $n_e^{-1}$ and variance $(\alpha_w n_e^2)^{-1}$.
Hence a large value of $\alpha_w$ results in a prior that is concentrated around $n_e^{-1}$ with a small variance.
If $n_e$ is large, this results in that only a few components are a priori given a large mixture weight.
This intuition is validated empirically by \cite{MalsinerFruhwirthGrun2016} which argue that this approach with $\alpha_w = 10$ usually recovers a mixture with the correct number of components.
We have also seen evidence supporting this in our preliminary work (not presented here).
Note that the consistency results obtained by \cite{RousseauMengersen2011} also holds in this case and this is the reason for choice of \eqref{eq:sparse:weights:prior} as the prior structure for the mixture weights in this paper.

\section{Hamiltonian Monte Carlo Computation of Posteriors}
\label{sec:hmc}
The parameter vector for the BARX model is given by
\begin{align}
	\theta
	=
	\{
	a_{1:n_a},
	b_{1:n_b},
	w_{1:n_e},
	\mu_{1:n_e},
	\sigma_{1:n_e},
	\sigma_{\mu},
	e_0
	\},
	\label{eq:parameter:vector}
\end{align}
which needs to be estimated from the input-output data.
Having specified the priors for these parameters, we now turn to the question of how to combine these priors with the likelihood of the observed data to deliver the required Bayesian posterior estimates.

We approach this via a computational Bayesian approach wherein a Markov chain is constructed with an invariant density which will converge to being equal to the posterior of interest.
That is, we construct a random number generator that generates realisations from the posterior, which can be used to obtain estimates of e.g, the posterior mean and its uncertainty.

This can be done using a Metropolis-Hastings (MH; \citealp{RobertCasella2004}) algorithm which operates by sampling from some \emph{proposal} Markov chain governed by
\begin{align}
	\theta'
	\sim
	q(\theta' | \theta_{k-1}),
\end{align}
where $q(\cdot)$ denotes some Markov kernel.
The candidate process that generates $\theta'$ is then modulated by setting $\theta_k \leftarrow \theta'$ with probability
\begin{align}
	\alpha( \theta', \theta_{k-1})
	=
	\frac{
		\pi( \theta')
	}{
		\pi(\theta_{k-1})
	}
	\frac{
		q(\theta_{k-1} | \theta')
	}
	{
		q(\theta' | \theta_{k-1})
	},
\end{align}
otherwise the candidate is rejected and $\theta_k \leftarrow \theta_{k-1}$.
Here, $\pi(\cdot)$ denotes the \emph{target distribution} which in our case is which is the parameter posterior
\begin{align}
	\pi(\theta)
	=
	p(\theta | y_{1:T})
	\propto
	p(\theta)
	p(y_{1:T} | \theta).
\end{align}

MH is very simple algorithm, but unfortunately its effectiveness is very much dependent on how well the proposal density $q(\cdot)$ is \emph{tuned} to the the target density $\pi(\cdot)$.
If the two are not similar then the acceptance probability $\alpha$ either rejects a great majority of candidates $\theta'$ that propose any significant movement, or accepts a high proportion of tiny movements, and in both cases generates highly correlated realisations with very poor sample average convergence rates to underlying true estimates.

HMC \citep{Neal2010,Betancourt2017} methods address this problem by replacing the target
density $\pi(\theta)$ with an extended target given by
\begin{align}
	H(\theta, p)
	=
	- \log \pi(\theta)
	+
	\frac{1}{2}
	p^{\top}
	M^{-1}
	p,
	\label{eq:hamiltonian}
\end{align}
which is known as the Hamiltonian in physics.
Here, we introduce an auxiliary \emph{momentum} variable $p$ to facilitate computation, which later can be removed by marginalisation.
A standard choice is that $p$ is a zero-mean Gaussian with the so-called mass matrix $M$ as its covariance matrix.

Samples from the posterior can be obtained by simulating the dynamical system described by \eqref{eq:hamiltonian}.
That is, we explore level sets of the posterior where the Hamiltonian is constant, which enables almost uncorrelated sampling from it compared with the small local moves usually employed within MH.
The possibility of larger steps using the Hamiltonian dynamics is what decreases the correlation in the samples, which is beneficial for the performance of the sampling.
The simulation is carried out by solving
\begin{subequations}
\begin{align}
	\frac
	{\partial \theta}
	{\partial t}
	&=
	\frac
	{\partial H(\theta, p)}
	{\partial p}
	=
	M^{-1}
	p,
	\\
	\frac
	{\partial p}
	{\partial t}
	&=
	-
	\frac
	{\partial H(\theta, p)}
	{\partial \theta}
	=
	\nabla_{\theta} \log \pi(\theta),
\end{align}%
\label{eq:hamiltonian:dynamics}%
\end{subequations}%
over a user chosen time period.
Note that the gradient of the log-posterior enters into the time derivative of the momentum.
Hence, it will help to guide the Markov process to areas of high posterior probability.

Unfortunately, it is not possible to solve \eqref{eq:hamiltonian:dynamics} in closed-form for this particular problem.
Instead, it is possible to make use of numerical integration methods to simulate the simulation.
However, not all methods are applicable for this, see \cite{Neal2010} for details, but leap-frog integrators can be employed to numerically simulate \eqref{eq:hamiltonian:dynamics}.
The resulting move from $\theta_{k-1}$ to $\theta'$ is accepted with probability
\begin{align}
	\alpha(\theta_k | \theta_{k-1})
	=
	\exp
	\Big(
	-
	H(\theta', p')
	+
	H(\theta_{k-1}, p_{k-1})
	\Big).
\end{align}
This basically accepts candidates if the Hamiltonian does not decrease significantly over the simulation.
Hence it corrects for the error introduced by the numerical integration.

The implementation of HMC requires the user to determine the mass matrix $M$ as well as the step-length and number of steps in the leap-frog algorithm.
In this paper, we make use of STAN~\citep{Stan2017} for implementing HMC and this software includes an adaptive method known as NUTS~\citep{HoffmanGelman2014} to set these parameters automatically.


\section{Numerical illustrations}
\label{sec:results}
We provide three different numerical illustrations to investigate the properties and the performance of the proposed BARX model using both synthetic and real-world data.
In each of the illustrations, the \textit{same} model structure and algorithmic settings are used and the \textit{only} difference is the input-output data supplied to the algorithm.
All implementation details are summarised in Appendix~\ref{app:impdetails} and the source code is available via GitHub, see Section~\ref{sec:conclusions}.

\begin{figure}[p]
	\centering
	\includegraphics[width=0.75\textwidth]{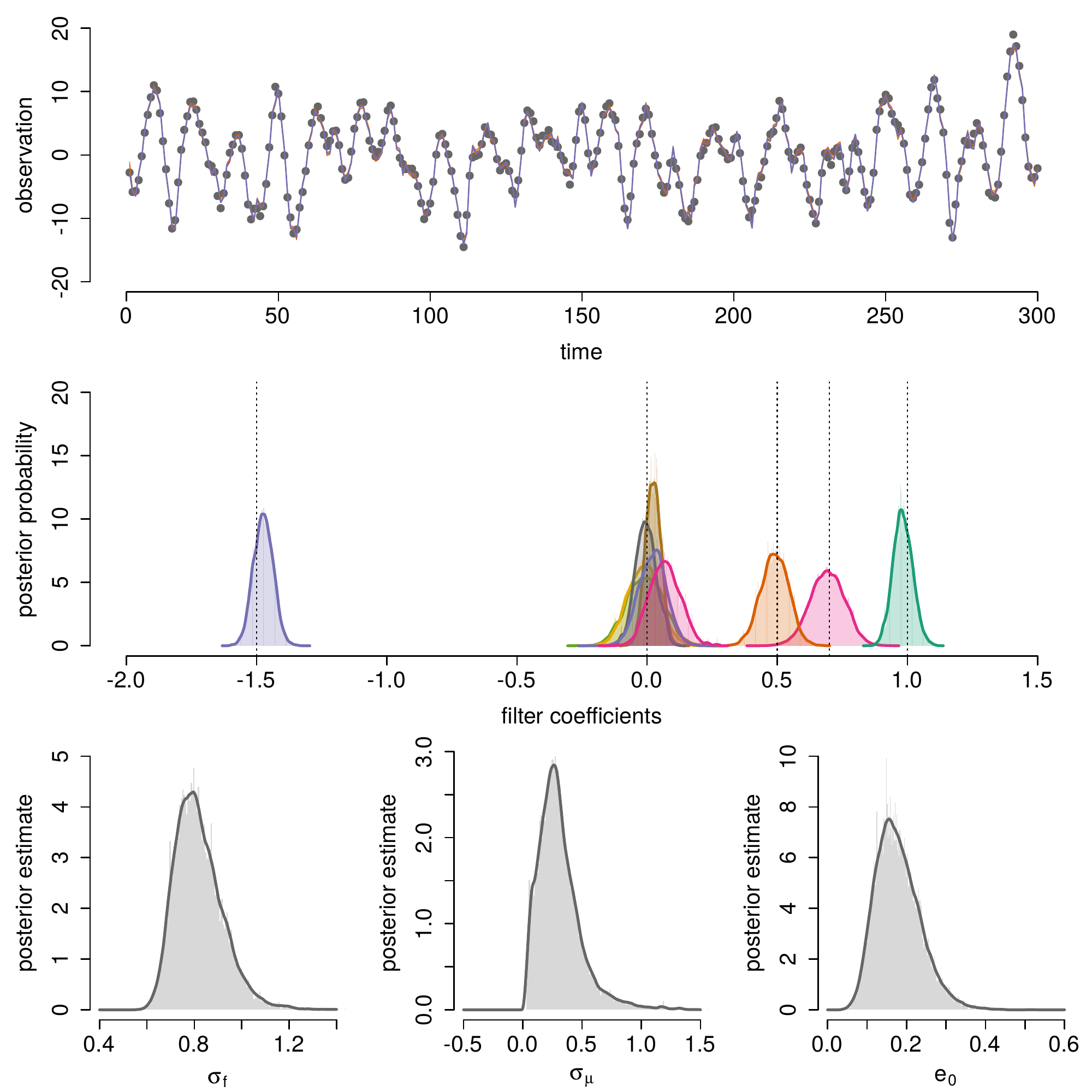}
	\caption{Top: the one-step-ahead predictors for \texttt{arx} (purple) and BARX (orange) models versus validation data (dots) in a model with unimodal Gaussian noise.
	The estimated posteriors of the system polynomial coefficients (middle) and priors (bottom) are also presented.
	Dotted vertical lines indicate the true system coefficients.}
	\label{fig:example1-arx-paper}
\end{figure}

\subsection{Synthetic data with Gaussian noise}
\label{sec:results:example1}
We generate a data set with $T=1,000$ observations using the system polynomial coefficients
\begin{align*}
	a_{1:2} = \{-1.5, 0.7\}, \quad b_{1:3} = \{0, 1, 0.5\},
\end{align*}
using a pseudo-binary input signal and adding standard Gaussian noise
to the output signal.
We partition the output signal into an estimation and a validation set, using $2/3$ and $1/3$ of the data, respectively.

The BARX estimate model is compared with the results of the \texttt{arx} command in MATLAB combined with a cross-validation scheme to find the model order.
Regarding the latter, we partition the estimation data into two sets of equal length and make use of the first part for estimating models where $n_a$ and $n_b$ varies between $1$ and $5$.
The model order is selected as the one that minimises the squared prediction error computed on the second part.
This approach selects $n_a=3$ and $n_b=5$.

Figure~\ref{fig:example1-arx-paper} summarises the results obtained from the experiment.
In the top plot, the validation data set (dots) is presented together with the one-step-ahead predictors obtained via the BARX model (orange) and \texttt{arx} (purple).
We note that the predictors are virtually the same for most time steps.
Using the predictors, the model fit is computed on validation data by
\begin{align*}
	\mathsf{MF}
	=
	100
	\left(
	1
	-
	\frac
	{\sum_{t=1}^{T_v} (y_t - \widehat{y}_t)^2}
	{\sum_{t=1}^{T_v} (y_t - \bar{y})^2}
	\right),
\end{align*}
where $\widehat{y}_t$ and $\bar{y}$ denote the one-step-ahead predictor and the sample mean of the output signal, respectively.
Note that the mean of the predictive distribution from BARX is used for this computation.
We obtain the model fit $96.6\%$ and $96.5\%$ for \texttt{arx} and BARX, respectively.
This indicates that BARX can replicate the result of \texttt{arx} for models with unimodal Gaussian noise and thereby offers a good sanity.
Here, \texttt{arx} runs in a few seconds and the estimation of BARX takes a few minutes.

The middle and bottom plots in Figure~\ref{fig:example1-arx-paper} present the estimates of the system coefficients and the estimates of the prior distributions for the BARX model, respectively.
Note that the sparsity of the horseshoe prior in the system coefficients has shrunk most of system coefficient posteriors towards zero.
At least four coefficients do not overlap zero and therefore remain statistically significant, which corresponds well to the model from which the data was generated.

From the posteriors of the priors, we can see that the prior for the system coefficients can accommodate larger deviations from zero than the prior for the mixture means.
This is reasonable as some of the system coefficients are quite far from zero.
Finally, we note that the prior on the mixture weights indicates a sparse behaviour as it is quite small and therefore promotes a few mixture components to have large weights with the rest being small.

\begin{figure}[p]
	\centering
	\includegraphics[width=0.75\textwidth]{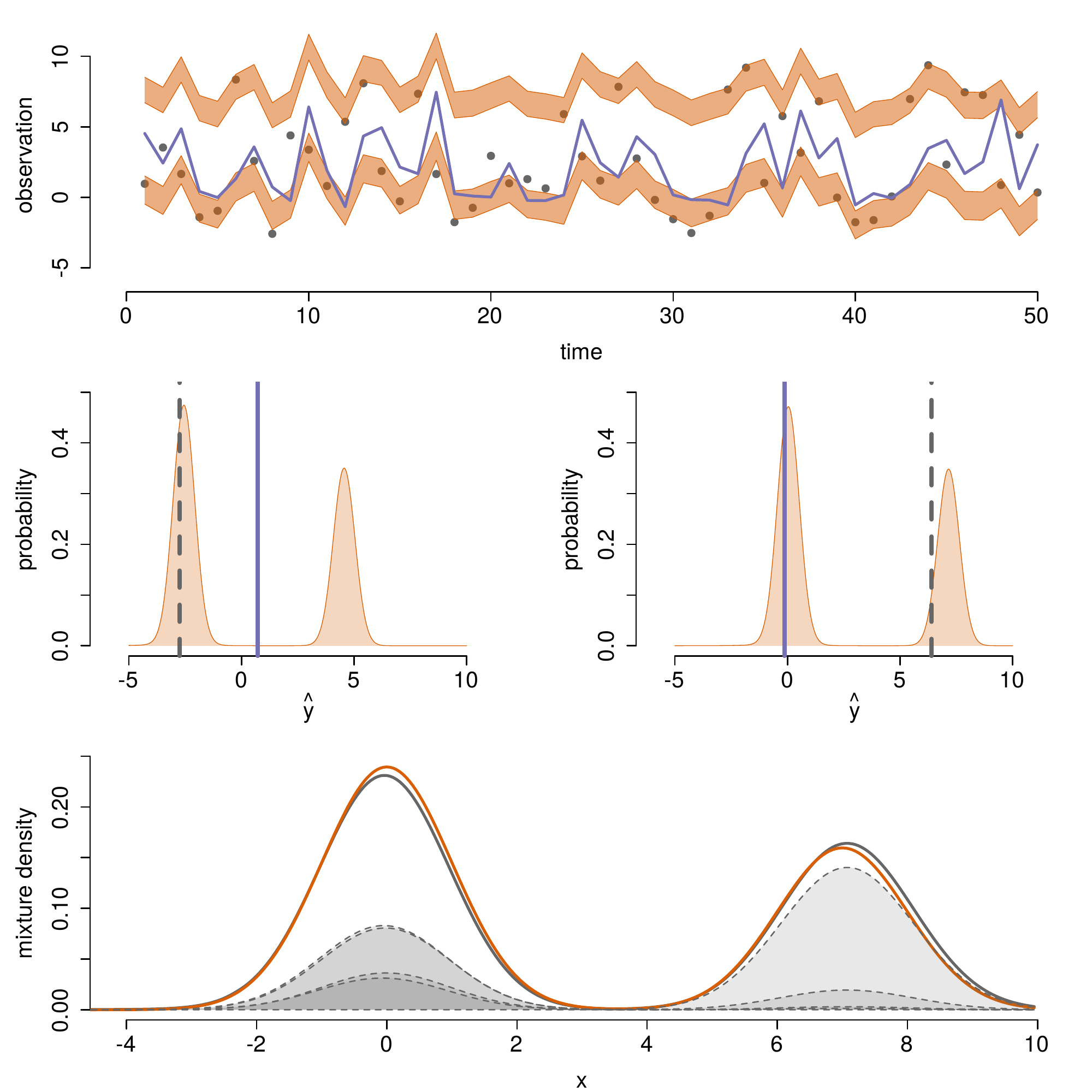}
	\caption{
		BARX with Gaussian mixture noise.
		Top: validation data (grey dots) with HPD of the one-step-head prediction distribution from BARX (orange) and the corresponding prediction from \texttt{arx} (purple).
		Middle: The one-step-head predictions for \texttt{arx} (solid line) and BARX (orange) at two time-steps with the true observations (dashed line).
		Bottom: the estimated noise distribution (grey solid line) together with the true distribution (orange) and the GMM components (grey areas).
		See the main text for details and a discussion.
		}
	\label{fig:example2-arxgmm-paper}
\end{figure}

\subsection{Synthetic data with Gaussian mixture noise}
\label{sec:results:example2}
We continue with a more interesting example where the output is corrupted with noise from a GMM with a multi-modal behaviour.
This might occur in practice due to unmodelled system behaviour of sensor imperfections.
We generate a data set with $T=1000$ observations from an ARX system \eqref{eq:arx} with system coefficients \eqref{eq:arx:polynomials}
\begin{align*}
	a_{2:3}
	=
	\{
		-0.25, 0.2
	\},
	\quad
	b_{1:3}
	=
	\{
		0, 1, 0.5
	\},
\end{align*}
and being driven by an exogenous input $u_t$ being an i.i.d. zero mean and unit
variance Gaussian process.
The observations $y_t$ involve a noise sequence $e_t$ which are i.i.d. realisations from a GMM (\ref{eq:noisemodel}) given by
\begin{align*}
	p(e_t)
	=
	0.4
	\cdot
	\mathcal{N}(e_t: 7, 1)
	+
	0.6
	\cdot
	\mathcal{N}(e_t: 0, 1).
\end{align*}
This means that the mean of the noise switches between $0$ and $7$ due to some unknown underlying process.
The model fit for this data set obtained by \texttt{arx} using the same approach as in Section~\ref{sec:results:example1} is $-0.22\%$.

This poor performance is due to the obvious violation of the Gaussian noise assumption, but it can be challenging to validate these assumptions in practice.
To compare, we fit the BARX model to the same data to see how its data-driven properties handle the bi-modal noise distribution.

Figure~\ref{fig:example2-arxgmm-paper} presents the results from obtained by BARX.
In the top part, we see the high posterior density (HPD) intervals of the one-step-ahead prediction distribution obtained from BARX (orange) together the \texttt{arx} predictor (purple) and the validation data (dots).
Note that the predictor from \texttt{arx} tries to capture the bi-modality by adjusting the mean but this results in poor performance as no data is found in the gap between the two modes.

In the middle, we present the one-step-ahead prediction distributions (orange) from BARX together from the same prediction from \texttt{arx} (purple solid line) and the true observation (grey solid line).
The \texttt{arx} predictor struggles in these to cases and is far away from the true observation which is covered by the distribution from BARX.
In the lower part, we present the true noise distribution (orange line) together with the BARX estimate (gray line). BARX is able to provide a good estimate of the noise distribution in this case.

As BARX is a Bayesian approach, the one-step-ahead predictor is actually a distribution which is able to capture the multi-modal nature of the data generating model.
This highlights the benefit of working with distributions of quantities instead of point-estimates, which are the standard approach in likelihood-based System Identification methods.
Hence, a distribution of the predictor could be useful within e.g., stochastic MPC.

\subsection{Real-world EEG data}
\label{sec:results:example3}
Finally, we revisit a real-world EEG data set analysed in \cite{DahlinLindstenSchonWills2012}, where the authors assume an ARX model with Student's $t$-distributed noise.
We apply the same procedure as before to estimate the BARX model.
There is no input signal in this data set.

\begin{figure}[p]
	\centering
	\includegraphics[width=0.75\textwidth]{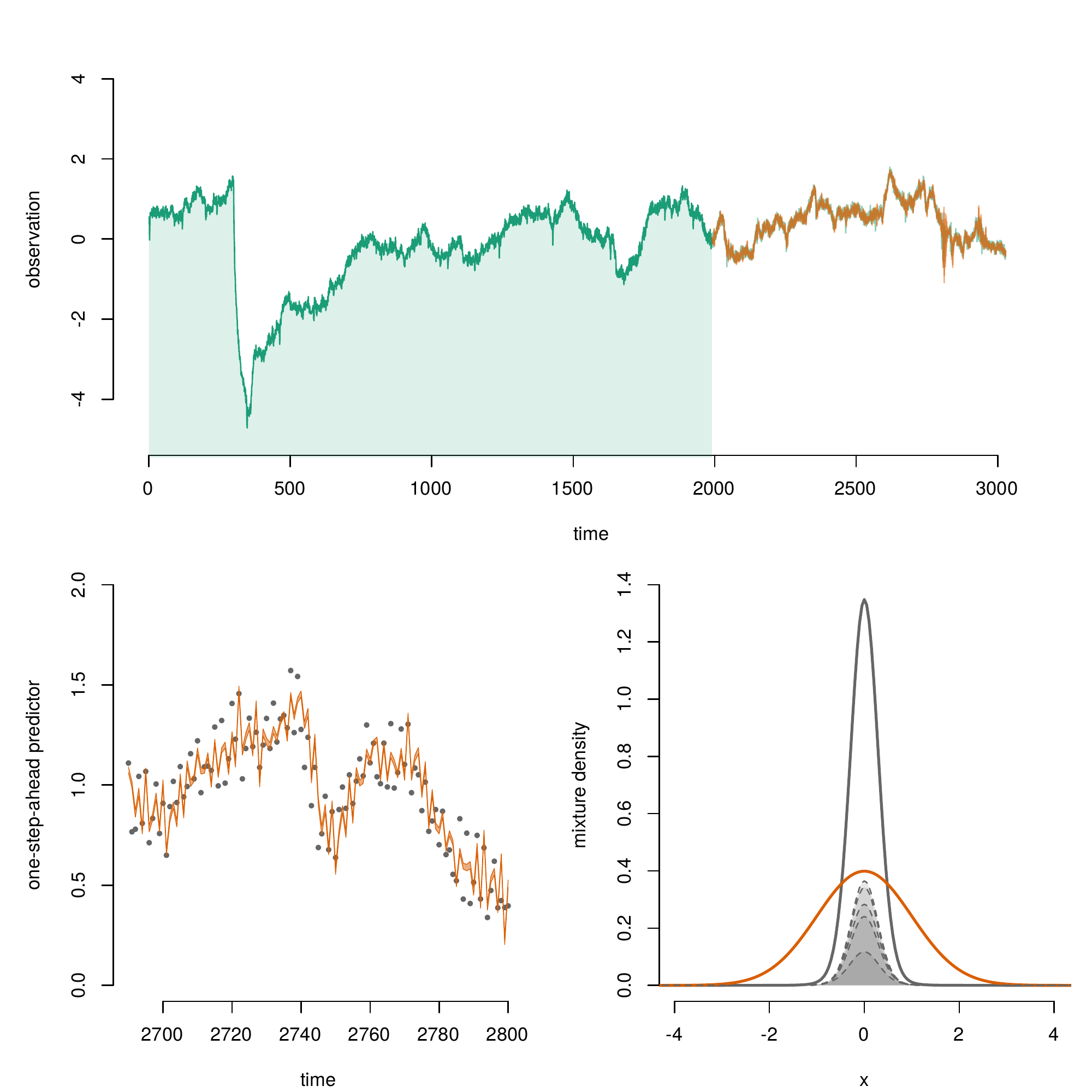}
	\caption{
		BARX model estimated on real-world EEG data.
		Top: the estimation data (shaded green) and validation data set (unshaded green) are presented together with the one-step-ahead predictor (orange).
		Bottom left: zoom-in of top plot with validation data indicated by dots.
		Bottom right: the estimate of the noise distribution with the best Gaussian approximation (orange) to the estimation data.}
	\label{fig:example3-eegdata}
\end{figure}

Figure~\ref{fig:example3-eegdata} presents the results obtained by using BARX.
The one-step-ahead predictors seem to perform well and this is reflected in a high model fit of $92.3\%$ on validation data for the BARX model, which is a substantial increase from the model fit of $85.6\%$ obtained in \cite{DahlinLindstenSchonWills2012}.

Moreover, we note that the estimated noise of the data set is clearly not Gaussian, as the behaviour of the tails are very different.
An interesting finding from this experiment is that only $\{a_1, a_2, a_4, a_5, a_6, a_7\}$ seems to be non-zero and contribute to the predictor.

\section{Conclusions}
\label{sec:conclusions}
The numerical illustrations indicate proposed BARX model can be useful in problems where non-standard measurement noise corruptions cause traditional system estimation approaches to fail by a significant margin.
BARX performs equivalently to standard methods when the measurement
noise is unimodal and Gaussian, which makes it an attractive option when the noise distribution is known as possibly multi-modal.

Hence, BARX is an automated data-driven approach that adapts its complexity to capture the dynamic behaviour observed in the data.
This is similar to many modern non-parameteric Bayesian models employed within e.g., machine learning, see \cite{Ghahramani2015}.
The main benefit with this kind of model is that its complexity is allowed to scale with the amount of available data.
It is therefore able to capture more and more complicated system behaviours as more data is accumulated.

Future work includes the generalisation to more flexible model structures such as Box-Jenkins transfer function models. Moreover, it would also be interesting to integrate this inference approach within MPC to create controllers that can handle e.g., multi-modal or heavy-tailed noise.

The source code and data used in this paper are available from GitHub \url{https://github.com/compops/barx-sysid2018/} and via Docker (see \texttt{README.md}).

\section*{Acknowledgements}
The EEG data was kindly provided by Eline Borch Petersen and Thomas Lunner at Eriksholm Research Centre, Oticon A/S, Denmark.
The authors would like to thank Johan W{\aa}gberg for comments on an earlier version of this paper.

\bibliographystyle{plainnat}
\bibliography{dahlin}

\appendix
\section{Implementation details}
\label{app:impdetails}
The implementation details are almost the same for all three illustrations in Section~\ref{sec:results}.
In the first two experiments, we make use of a maximum order of $5$ for the system polynomials, i.e., $n_a=n_b=5$ and use $n_e=5$ components for the noise distribution mixture.
In the third experiment, we set $n_a=10$ and $n_b=0$ (as no input is present).
These values have not been calibrated in any way and were just selected arbitrarily.
The hyperpriors are selected as in Section~\ref{sec:modelling}.

The HMC sampler is implemented using the STAN software \citep{Stan2017} and NUTS \citep{HoffmanGelman2014} to adaptive the settings of the algorithm.
We run the sampler for $30,000$ iterations discarding the first $15,000$ iterations as burn-in.

The initialisation of the system coefficients and the means of the mixture components greatly influences the accuracy of the estimates.
We therefore initialise the former randomly over the interval $(-1, 1)$ and the latter on an equally spaced grid over the range of the output data.
All mixture weights are initialised as $1/n_e$ and the remaining parameters are initialised as $1$.

\end{document}